\documentclass[prb,superscriptaddress]{revtex4}            
\usepackage{amsmath}
\usepackage{amsfonts}
\usepackage[dvips]{graphicx}
\usepackage{epsfig}
\newcommand{\eref}[1]{(\ref{#1})}
\newcommand{\sref}[2]{(\ref{#1}-\ref{#2})}
\newcommand{\esref}[3]{(\ref{#1},\ref{#2}-\ref{#3})}
\newcommand{\tref}[2]{(\ref{#1},\ref{#2})}
\newcommand{\Eref}[1]{Eq.~(\ref{#1})}
\newcommand{\Mref}[2]{Eqs.~(\ref{#1}-\ref{#2})}
\newcommand{\Tref}[2]{Eqs.~(\ref{#1},\ref{#2})}
\newcommand{\Fref}[1]{Fig.~\ref{#1}}
\newcommand{\Sref}[1]{Section ~\ref{#1}}

\newcommand{\rmi}{\mathrm i}

\newcommand{\e}{\exp}
\newcommand{\rmd}{\mathrm d}
\renewcommand{\vec}{\boldsymbol}

\newcommand{\citeo}[1]{Ref.~\cite{#1}}

\newcommand{\degre}{^{\mathrm o}}
\newcommand{\ep}{\epsilon}
\newcommand{\x}{\vec{x}}
\newcommand{\E}{\vec{E}}
\newcommand{\p}[1]{\vec{p}_{#1}}

\newcommand{\uu}[1]{\vec{u}_{#1}}
\newcommand{\op}[1]{\overline{\vec{#1}}}
\newcommand{\hux}{h_1(\vec{x})}
\newcommand{\hdx}{h_2(\vec{x})}

\newcommand{\hvec}[1]{\hat{\vec{#1}}}
\newcommand{\ehp}[1]{\hvec{e}_H(\p{#1})}
\newcommand{\evp}[2]{\hvec{e}_V^{#1}(\p{#2})}

\newcommand{\ez}{\hvec{e}_z}

\newcommand{\intp}[1]{\int \frac{\rmd^2 \vec{p}_{#1}}{(2\pi)^2}}
\newcommand{\intu}[1]{\int \frac{\rmd^2 \vec{u}_{#1}}{(2\pi)^2}}
\newcommand{\iintp}[2]{\int \!\!\! \int \frac{\rmd^2\vec{p}_{#1}}{(2\pi)^2} \frac{\rmd^2 \vec{p}_{#2}}{(2\pi)^2}}

\newcommand{\ga}{\gamma}
\newcommand{\alp}[2]{\alpha_{#1}(\vec{p}_{#2})}
\newcommand{\alu}[2]{\alpha_{#1}(\vec{u}_{#2})}
\newcommand{\Rp}[2]{\op{R}(\p{#1}|\p{#2})}
\newcommand{\no}{\nonumber}

\newenvironment{matrice}{\begin{pmatrix}}{\end{pmatrix}}

\begin{document}
\title{Backscattering enhancement of an electromagnetic wave scattered \\
by two-dimensional rough layers}
\author{Antoine Soubret}
\affiliation{Thales Optronique, 
BP 55, 78233 Guyancourt Cedex, France.}
\affiliation{Centre de Physique Th\'eorique, CNRS-Luminy Case
907, 13288 Marseille Cedex 9, France.}
\author{G\'erard Berginc}
\affiliation{Thales Optronique, 
BP 55, 78233 Guyancourt Cedex, France.}
\author{Claude  Bourrely}
\affiliation{Centre de Physique Th\'eorique, CNRS-Luminy Case
907, 13288 Marseille Cedex 9, France.}
\date{December 2000}

\begin{abstract}
The problem of an electromagnetic wave scattering by a slab with two
rough boundaries is solved by a small-perturbation method under the
Rayleigh hypothesis. In order to obtain a perturbative development, we use
a systematic procedure which involves integral equations called the reduced
Rayleigh equations. Then we will show for a dielectric slab deposited
on a silver film that the  backscattering enhancement can be produced 
by guided waves which interact with the two rough surfaces.
\end{abstract}

\pacs{PACS number(s): 42.25.Fx, 42.25.Ja, 42.25.Gy, 78.66.Bz, 78.20.e, 73.20.Mf}
\maketitle    

\section{Introduction}
Scattering of electromagnetic waves from multilayer structures is a
phenomenon which is of interest in many area of physics such as 
remote sensing or optical industry,  where for example metallic
surfaces have a dielectric coating. An  extended review of experimental
and theoretical works on optical multilayers can be found in \cite{Oli},
and references on recent papers are given in \cite{Fuks}. Many works 
\cite{Bousquet,Elson,Amra,Fuks}
deal with the small perturbation method to investigate the behaviour 
of multilayer structure. However, due to calculations complexity, the
analytical results are given only to first order in the rms-height of
each rough boundary, thus the interaction between the rough surfaces
can only be taken into account if  the rough surfaces 
are correlated \cite{Elson} in the mean procedure. Our purpose is  to show
how light can interact with several rough surfaces, to this end, we
have choosen the most simple system depicted in \Fref{systeme0}, where
we have three regions with different permitivities separated by two rough
surfaces. The calculations have been made under the hypothesis of the
small perturbation method, initially developped by
Rice\cite{Rice,Voro}. Due to the great complexity to derive high orders
of the perturbative development, Rice's original method is rather
difficult to apply. One way to overcome these difficulties is to use 
integral equations called the reduced Rayleigh equations, they were first 
obtained by
Celli {\it et al} \cite{Celli}, and later have been generalized in
\citeo{Soubret} to take into account upward fields in the slab
medium. When  combining these equations
we obtain an integral equation where only the scattering matrix of the 
whole structure has to be determined, next  we have developed a
systematic method to calculate the perturbative development.
We have shown by means of numerical simulations, how different mechanisms
responsible of the enhanced backscattering occur. This phenomenon which
has been predicted \cite{Mara1,Mara2,Mara3,Mendez} and
observed \cite{West} in the case of a random metal surface, manifests
itself as a well-defined peak in the retroreflection direction through the
angular dependence of the intensity related to the diffuse component of the
scattered field. For a metallic surface, the phenomenon is produced by
the interference of waves which excite surface plasmon polariton along a certain
path and
then follow the same path but in the reverse direction. For a dielectric bounded
by metallic plate 
with one rough surface \cite{Mara}, the enhanced backscattering is produced by
a similar mechanism where the surface plasmon polariton is replaced 
by a slab guided wave. In this case the incident wave  excites in a first time 
a guided mode due to surface roughness, and then the 
roughness transforms the surface wave into a bulk wave. 
Furthermore, if the slab supports several guided modes, recent
investigations \cite{Mara,Soubret}  have shown the presence of
additionnal peaks, called {\it satellite peaks}, in the angular distribution 
of the incoherent intensity. When the slab has two rough boundaries,
the enhancement of backscattering can be produced by two kinds of
interaction. The classic one, where the wave is scattered two times
by the same boundary, and a new one (see \Fref{systeme0}) where the
wave is transmitted at a point $A$ of the surface  without being diffused, then
the wave is scattered by the second rough
surface at a point $B$ and by the first one in $C$. If the wave follows the same
path in the reverse direction and excites the same guided mode, then
phase difference between the two path is
$\Delta\phi=\vec{r}_{BC}\cdot(\vec{k}+\vec{k}_i)$, where
$\vec{r}_{BC}$ is the distance between $B$ and $C$. Thus, in the
anti-specular direction ($\vec{k}=-\vec{k}_i$), this phase difference
is independent of the random position of $B$ and $C$ which produce the 
backscattering peak.

The outline of the paper is as follows. In {\Sref{Random}} we present the 
system under study and a description of the rough
surfaces statistics.
In {\Sref{Scattering}} we define the plane wave representation of the
electric field in the polarization basis and also the
scattering matrix. {\Sref{Diffuse}} is devoted to the calculation of the 
incoherent cross-section as a function of the perturbative
development. In {\Sref{Perturbative}} we derive an
integral equation from the reduced Rayleigh equations, which provides a
systematic way to obtain the perturbative development.
The resulting expressions are numerically computed in the case of a
dielectric slab deposited on a  rough silver surface in {\Sref{Numerical}}.
The calculated incoherent intensity will show a narrow enhanced backscattering. 
Conclusions drawn from our  results are presented
and discussed in {\Sref{Conclusions}}. In an Appendix, we collect
matrices used in  the calculations. 

\section{The random surface}
\label{Random}
\begin{figure}[ht]
\input{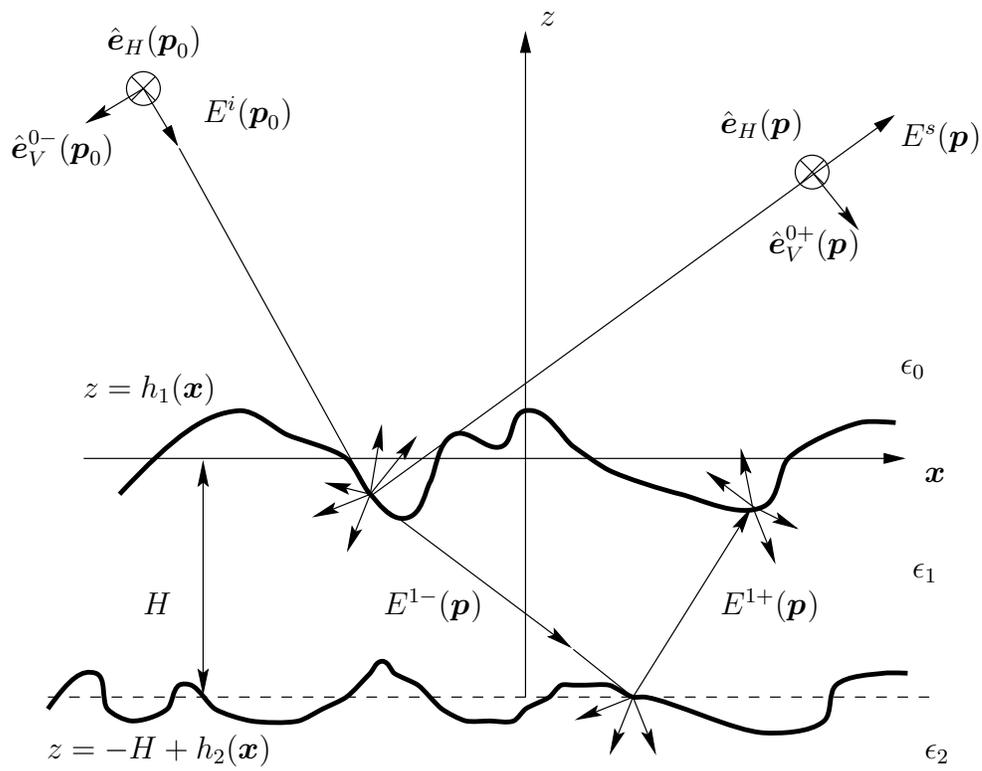}
\caption{A rough surface with an incident wave coming from the medium
  0 and scattered by a slab with two rough surfaces.}
\label{systeme}
\end{figure}

The system which is considered in this paper is depicted in
\Fref{systeme}.

The three regions  are characterized by an isotropic,
homogeous dielectric constant $\ep_0$, $\ep_1$ and $\ep_2$ respectively. 
The two boundaries are located at $z=h_1(\x)$ and 
$z=-H+h_2(\x)$, $\x=(x,y)$,
and  these three media are separated by rough
surfaces described statistically. In fact, we assume that
$h_1(\x)$ and $h_2(\x)$ are stationary, isotropic uncorrelated
Gaussian random processes defined by their moments :

\begin{align}
<h_i(\x)>&=0\,  ,\label{mom1}\\
<h_i(\x)\,h_i(\vec{x'})>&=W_i(\vec{x}-\vec{x'})\, ,\\
<h_1(\x)\,h_2(\vec{x'})>&=0 \label{mom4}
\end{align}
where $i=1,2$, and the angle brackets denote an average over the ensemble of
realizations of the function $\hux$ and $\hdx$.
In this work we will use a Gaussian form for the surface-height correlation 
function $W_1(\x)$ and $W_2(\x)$:
\begin{align}
W_i(\x)=\sigma_i^2\,\e(-\x^2/l_i^2)\, ,
\end{align}
where $\sigma_{i}$ is the rms height of the surface $h_i(\x)$, and $l_{i}$ is 
the transverse correlation length. In momentum space we have:
\begin{align}
<h_i(\p{})>&=0\,,\\
<h_i(\p{})\,h_i(\vec{p}')>
&=(2\pi)^2\,\delta(\p{}+\vec{p}')\,W_i(\p{})\, ,\label{momp2}\\
<h_1(\p{})\,h_2(\vec{p}')>&=0\, .
\end{align}
with 
\begin{align}
W_i(\p{})&\equiv\int\,\rmd^2\x\,W_i(\x)\,\e(-i\p{}\cdot\x)\\
&=\pi\,\sigma_i^2\,l_i^2\,\e(-\p{}^2\,l_i^2/4)\, .
\end{align}

\section{The scattering matrix}
\label{Scattering}
We suppose that the slab is illuminated from the media $0$ by an 
electromagnetic wave of 
pulsation $\omega$. In the following we will omit  the time dependence
$\e(-\rmi\,\omega\,t)$.
The field $\E^{0}$ in the media $0$ can be written as a superposition of an 
incident and scattered fields :
\begin{equation}
\vec{E}^{0}(\x,z)=\vec{E}^{i}(\p{0})\,\e(\rmi\p{0}\cdot\x-\rmi\alp{0}{0}\,z) 
+\intp{} \vec{E}^{s}(\p{})\,\e(\rmi\p{}\cdot\x+\rmi\alp{0}{}\,z)
\label{DefEp}\, .
\end{equation}
where 
\begin{align}
\alp{0}{}&\equiv(\ep_0\,K_0^2-\p{}^2)^{\frac{1}{2}}\, , \\
\E^{i}(\p{0})&=E^i_{V}(\p{0})\,\evp{0-}{0}+E^i_{H}(\p{0})\,\ehp{0}\,
\\
\E^{s}(\p{})&=E^s_{V}(\p{})\,\evp{0+}{}+E^s_{H}(\p{})\,\ehp{}\,.
\end{align}
The subscript $H$ refers to the horizontal polarization $(TE)$ and $V$
to the vertical polarization $(TM)$, and are defined by the two vectors:
\begin{align}\ehp{}& =\ez \times
\hvec{p} \, ,
\label{TE0}\\
\evp{0\pm}{} & =\pm
\frac{\alpha_0(\vec p)}{\sqrt{\ep_0}K_0}\hvec{p}-\frac{||\vec p||}
{\sqrt{\ep_0}K_0}\ez \, ,
\label{TM0}
\end{align}
where the minus sign refers to incident wave and the plus sign to the 
scattered wave.
It has to be noticed that the vector $\E^{s}(\p{})$ and
$\E^{i}(\p{0})$ are expressed in different basis due the fact that
$\evp{0\pm}{}$
and $\evp{1\pm}{}$ depend on $\p{}$. 
In medium 1, we have a similar expression namely:
\begin{equation}
\vec{E}^{1}(\vec{r})=\intp{}
\vec{E}^{1-}(\vec{p})\e(\rmi\p{}\cdot\x-\rmi\,\alp{1}{}z)
+\intp{}\vec{E}^{1+}(\p{})\e(\rmi\p{}\cdot\x+\rmi\,\alp{1}{}z)\, ,
\label{DefE1}
\end{equation}
where
\begin{align}
\alp{1}{}\equiv & (\ep_1K_0^2-\p{}^2)^{\frac{1}{2}} \, .\label{alpha1}
\end{align}
The field $\E^{1-}$ is decomposed in the basis $(\evp{-}{},\ehp{})$, and 
$\E^{1+}$ in the basis $(\evp{+}{},\ehp{})$ with 
\begin{align}\ehp{}& =\ez \times
\hvec{p} \, ,
\label{TE1}\\
\evp{1\pm}{} & =\pm
\frac{\alpha_1(\vec p)}{\sqrt{\ep_1}K_0}\hvec{p}-\frac{||\vec p||}
{\sqrt{\ep_1}K_0}\ez \, .
\label{TM1}
\end{align}
We now introduce the definition the scattering matrix :
\begin{align}
\E^{s}(\p{})\equiv\op{R}(\p{}|\p{0})\cdot\E^{i}(\p{0})\, ,
\end{align}
where $\op{R}(\p{}|\p{0})$ is a two dimensional matrix which can be
written in the following form:
\begin{displaymath}
\Rp{}{0}=\begin{matrice} R_{VV}(\p{}|\p{0}) & R_{VH}(\p{}|\p{0}) \\
R_{HV}(\p{}|\p{0}) & R_{HH}(\p{}|\p{0})
\end{matrice} \, .
\end{displaymath}
\section{Diffuse cross-section}
\label{Diffuse}
\begin{figure}[ht]
\input{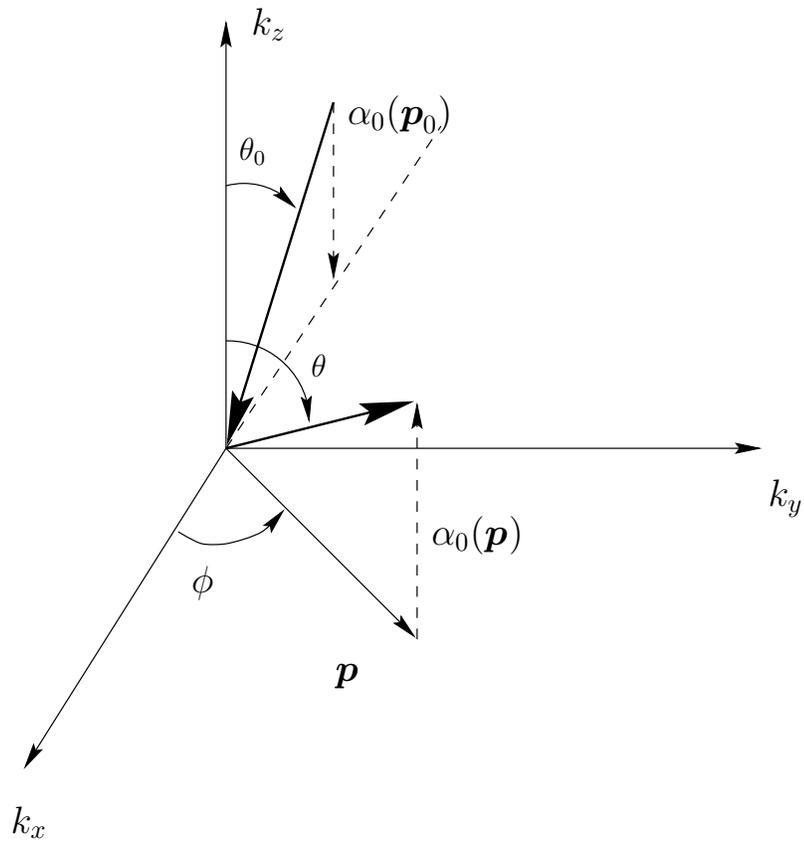}
\caption{Definition of the scattering vector.}
\label{vectonde}
\end{figure}

In a previous  work~\cite{Soubret} we have defined a new product 
$\odot$ between two dimensional matrices in the form :
\begin{align}
\op{f}\odot \op{g} &=  \begin{pmatrix} f_{VV} & f_{VH} \\f_{HV}
& f_{HH} \end{pmatrix} \odot \begin{pmatrix} g_{VV} & g_{VH}
\\g_{HV} & g_{HH} \end{pmatrix} \\
&\equiv \left(\begin{array}{cccc} 
f_{VV}g^*_{VV} & f_{VH} g_{VH}^* & Re(f_{VV}g_{VH}^*) & -Im(f_{VV}g_{VH}^*)\\
f_{HV}g^*_{HV} & f_{HH} g_{HH}^* & Re(f_{HV}g_{HH}^*) & -Im(f_{HV}g_{VH}^*)\\
2Re(f_{VV}g^*_{HV}) & 2Re(f_{VH} g_{HH}^*) & Re(f_{VV} g_{VV}^*+f_{HV} g_{VH}^*)
& -Im(f_{VV}g_{HH}^*-f_{VH}g_{HV}^*)\\
2Im(f_{VV}g^*_{HV}) & 2Im(f_{VH} g_{HH}^*) & Im(f_{VV} g_{VV}^*+f_{HV} 
g_{VH}^*)& 
Re(f_{VV}g_{HH}^*-f_{VH}g_{HV}^*)\end{array} \right) \, . \nonumber
\end{align}
which allows to write the incoherent  Muller matrix \cite{Ishi2} in a condensed 
expression:
\begin{align}
\op{M}^{incoh}(\p{}|\p{0})
&=\frac{K^2_0 \cos^2 \theta}{(2\pi)^2}\left[<\op{R}(\p{}|\p{0})
\odot\op{R}(\p{}|
\p{0})>-<\op{R}(\p{}|\p{0})>\odot<\op{R}(\p{}|\p{0})>\right]\, .
\label{incoh}
\end{align}
Furthermore, we define the generalization of the classical bistatic
coefficient \cite{Kong} also called mean differential
coefficient \cite{McGurn} by :
\begin{equation}
\op{\ga}^{incoh}(\p{}|\p{0})\equiv\frac{1}{A\,\cos\theta_0}
\op{M}^{incoh}(\p{}|\p{0})
\end{equation}
where
$A$ is the area of the illuminated surface and $\theta_0$ the incident
angle (see \Fref{vectonde}).
In this paper we are interested by the perturbative development of the 
scattered fields as a function of the surface elevations $h_1$ and
$h_2$. In a perturbative expansion of the scattering matrix, the
terms  which contain an expression like $h_1^n\,h_2^m$  will be denoted
$\op{R}^{(nm)}$, so that the perturbative development of $\op{R}$ becomes:
\begin{equation}
\op{R}=\op{R}^{(00)}+\op{R}^{(10)}+\op{R}^{(01)}+\op{R}^{(11)}+\op{R}^{(20)}
+\op{R}^{(21)}+\op{R}^{(12)}+\op{R}^{(22)}+\op{R}^{(30)}+\op{R}^{(03)}+\dots 
\label{dev}
\end{equation}
With this decomposition and using \Tref{mom1}{mom4} with the fact that 
$h_1$ and $h_2$ are Gaussian random processes {\it i.e.} 
\begin{align}
<h_1^{2p+1}(\x)>&=0\, ,\quad  p \, \mbox{a positive integer}
\\
<h_2^{2p+1}(\x)>&=0
\end{align}
the incoherent bistatic matrix will be given by the contribution of
three terms:
\begin{align}
\op{\ga}^{incoh}(\p{}|\p{0})=\op{\ga}_u^{incoh}(\p{}|\p{0})+
\op{\ga}_d^{incoh}(\p{}|\p{0})+\op{\ga}_{ud}^{incoh}(\p{}|\p{0})\, ,
\label{gaincoh}
\end{align}
where 
\begin{align}
\op{\ga}_u^{incoh}(\p{}|\p{0})=\frac{K^2_0 \cos^2
  \theta}{A\,(2\pi)^2\,\cos\theta_0}\left[<\op{R}^{(10)}\odot
  \op{R}^{(10)}>+<\op{R}^{(20)}\odot
  \op{R}^{(20)}>+<\op{R}^{(30)}\odot
  \op{R}^{(10)}>\right]\, ,\label{gau}
\end{align}
corresponds to the incoherent bistatic matrix for the slab where only
the upper surface has a roughness ($h_2(\x)=0$), and its expansion is made 
up to order four in the rms-height elevation $\sigma_1$.
Similarly,
\begin{align}
\op{\ga}_d^{incoh}(\p{}|\p{0})=\frac{K^2_0 \cos^2
  \theta}{A\,(2\pi)^2\,\cos\theta_0}\left[<\op{R}^{(01)}\odot
  \op{R}^{(01)}>+<\op{R}^{(02)}\odot
  \op{R}^{(02)}>+<\op{R}^{(03)}\odot
  \op{R}^{(01)}>\right]\, ,\label{gad}
\end{align}
is associated to a system where  only the bottom surface is rough
($h_1(\x)=0$), where also the perturbative development is made up to order four in
$\sigma_2$.
The last term $\op{\ga}_{ud}^{incoh}$ contains terms which describe
the  scattering process between the two rough surfaces  where only the leading 
terms are retained:
\begin{align}
\op{\ga}_{ud}^{incoh}(\p{}|\p{0})=\frac{K^2_0 \cos^2
  \theta}{A\,(2\pi)^2\,\cos\theta_0} & \left[ 
 <\op{R}^{(10)}\odot \op{R}^{(12)}>
+<\op{R}^{(12)}\odot \op{R}^{(10)}> \right. \no \\
& \,+<\op{R}^{(01)}\odot \op{R}^{(21)}>
+<\op{R}^{(21)}\odot \op{R}^{(01)}>\no \\
& \left.+<\op{R}^{(11)}\odot \op{R}^{(11)}>+\dots \right]\, ,\label{gaud}
\end{align}
More precisely  all terms up of order $\sigma_1^2\,\sigma_2^2$ are included. 
If the value of $\sigma_1$ and $\sigma_2$ are of the same order of magnitude, 
these terms will be comparable to the order four in the expressions
\tref{gau}{gad}. Thus we can suppose that the following terms in the
expansion \eref{gaud}, which are of order $\sigma_1^4\,\sigma_2^2$, 
$\sigma_1^2\,\sigma_2^4$, $\sigma_1^4\,\sigma_2^4$
will be negligible compared to those retained  in \Eref{gaud}.
However, due to the complexity of the perturbative
development, these terms of the sixth order have not been calculated.
In the following section, we will show how the
perturbative development can be put in the following form:
\begin{align}
\op{R}^{(10)}(\p{}|\p{0})&=\alp{0}{0}\,\op{X}^{(10)}(\p{}|
\p{0})\,h_1(\p{}-\p{0})\,,\label{R10}\\
\op{R}^{(01)}(\p{}|\p{0})&=\alp{0}{0}\,\op{X}^{(01)}(\p{}|
\p{0})\,h_2(\p{}-\p{0})\,,
\end{align}
\begin{align}
\op{R}^{(11)}(\p{}|\p{0})=\alp{0}{0}\,\intp{1}\,
&\left[\op{X}^{(11)12}(\p{}|\p{1}|\p{0})\,h_1(\p{}-\p{1})\,\,h_2(\p{}-\p{0})
\right.\no \\
&\left.+\op{X}^{(11)21}(\p{}|\p{1}|\p{0})\,h_2(\p{}-\p{1})\,\,
h_1(\p{}-\p{0})\right]\,,
\end{align}
\begin{align}
\op{R}^{(20)}(\p{}|\p{0})&=\alp{0}{0}\,\intp{1}\,
\op{X}^{(20)}(\p{}|\p{1}|\p{0})\,h_1(\p{}-\p{1})\,h_1(\p{1}-\p{0})\,,\\
\op{R}^{(02)}(\p{}|\p{0})&=\alp{0}{0}\,\intp{1}\,
\op{X}^{(02)}(\p{}|\p{1}|\p{0})\,h_2(\p{}-\p{1})\,h_2(\p{1}-\p{0})\,,
\end{align}
\begin{align}
\op{R}^{(21)}(\p{}|\p{0})=\alp{0}{0}\,\iintp{1}{2}\,
&\left[\op{X}^{(21)112}(\p{}|\p{1}|\p{2}|\p{0})\,h_1(\p{}-\p{1})\,
h_1(\p{1}-\p{2})\,h_2(\p{2}-\p{0})\right.\no \\
&+\op{X}^{(21)121}(\p{}|\p{1}|\p{2}|\p{0})\,h_1(\p{}-\p{1})\,
h_2(\p{1}-\p{2})\,\,h_1(\p{2}-\p{0})\,\no\\
&\left.+\op{X}^{(21)211}(\p{}|\p{1}|\p{2}|\p{0})\,h_2(\p{}-\p{1})\,
h_1(\p{1}-\p{2})\,\,h_1(\p{2}-\p{0})\right]\,,
\end{align}
\begin{align}
\op{R}^{(12)}(\p{}|\p{0})=\alp{0}{0}\,\iintp{1}{2}\,
&\left[\op{X}^{(12)221}(\p{}|\p{1}|\p{2}|\p{0})\,h_2(\p{}-\p{1})\,
h_2(\p{1}-\p{2})\,h_1(\p{2}-\p{0})\right.\no \\
&+\op{X}^{(12)212}(\p{}|\p{1}|\p{2}|\p{0})\,h_2(\p{}-\p{1})\,
h_1(\p{1}-\p{2})\,\,h_2(\p{2}-\p{0})\,\no\\
&\left.+\op{X}^{(12)122}(\p{}|\p{1}|\p{2}|\p{0})\,h_1(\p{}-\p{1})\,
h_2(\p{1}-\p{2})\,\,h_2(\p{2}-\p{0})\right]\,,
\end{align}
\begin{align}
\op{R}^{(30)}(\p{}|\p{0})&=\alp{0}{0}\,\iintp{1}{2}\,
\op{X}^{(30)}(\p{}|\p{1}|\p{2}|\p{0})\,h_1(\p{}-\p{1})\,h_1(\p{1}-\p{2})\,
h_1(\p{2}-\p{0})\,,\\
\op{R}^{(03)}(\p{}|\p{0})&=\alp{0}{0}\,\iintp{1}{2}\,
\op{X}^{(03)}(\p{}|\p{1}|\p{2}|\p{0})\,h_2(\p{}-\p{1})\,h_2(\p{1}-\p{2})\,
h_2(\p{2}-\p{0})\,.\label{R03}
\end{align}
In these expressions we have added superscripts in some terms 
 to indicate the order of apparition of the functions
$h_1$ and $h_2$. For example, in $\op{X}^{(21)121}$ the superscript
$121$ indicates that it is the coefficient associated with the product
$h_1(\p{}-\p{1})\,h_2(\p{1}-\p{2})\,\,h_1(\p{2}-\p{0})$.
When we combine \Eref{momp2}, $\delta(0)=A/(2\pi)^2$ and the previous
development,
we obtain the following expression for the quantities \sref{gau}{gaud}:
\begin{align}
\op{\ga}_u^{incoh}(\p{}|\p{0})
&=\frac{K^4_0 \cos^2 \theta\,\cos
\theta_0}{(2\pi)^2}\,\left[\op{I}^{(10-10)}(\p{}|\p{0})+
\op{I}^{(20-20)}(\p{}|\p{0})+
\op{I}^{(30-10)}(\p{}|\p{0})\right]\, ,\label{incoh1}\\
\op{\ga}_d^{incoh}(\p{}|\p{0})
&=\frac{K^4_0 \cos^2 \theta\,\cos
\theta_0}{(2\pi)^2}\,\left[\op{I}^{(01-01)}(\p{}|\p{0})
+\op{I}^{(02-02)}(\p{}|\p{0})+
\op{I}^{(03-01)}(\p{}|\p{0})\right]\, ,\label{incoh2}\\
\op{\ga}_{ud}^{incoh}(\p{}|\p{0})
&=\frac{K^4_0 \cos^2 \theta\,\cos
\theta_0}{(2\pi)^2}\,\left[\op{I}^{(12-10)}(\p{}|\p{0})
+\op{I}^{(11-11)}(\p{}|\p{0})+
\op{I}^{(21-01)}(\p{}|\p{0})\right]\label{incoh3}\, ,
\end{align}
where
\begin{align}
\op{I}^{(10-10)}(\p{}|\p{0})&= W_1(\p{}-\p{0})\,\op{X}^{(10)}(\p{}|\p{0})
\odot\op{X}^{(10)}
(\p{}|\p{0})\label{ich10}\\ 
\op{I}^{(20-20)}(\p{}|\p{0})&= \intp{1}\,W_1(\p{}-\p{1})W_1(\p{1}-\p{0})\,
\op{X}^{(20)}(\p{}|\p{1}|\p{0})\no \\ &\odot
\left[\op{X}^{(20)}(\p{}|\p{1}|\p{0})+\op{X}^{(20)}(\p{}|\p{}+\p{0}-\p{1}|
\p{0})\right]\label{ich20}\\
\op{I}^{(30-10)}(\p{}|\p{0})&= W_1(\p{}-\p{0})\left[\op{X}^{(10)}(\p{}|\p{0})
\odot\op{X}^{(30)}(\p{}|\p{0})
+\op{X}^{(30)}(\p{}|\p{0})\odot
\op{X}^{(10)}(\p{}|\p{0})\right]\label{ich30}\, ,
\end{align}
\begin{align}
\op{I}^{(01-01)}(\p{}|\p{0})&= W_2(\p{}-\p{0})\,\op{X}^{(01)}(\p{}|\p{0})
\odot\op{X}^{(01)}
(\p{}|\p{0})\label{ich01}\\ 
\op{I}^{(02-02)}(\p{}|\p{0})&= \intp{1}\,W_2(\p{}-\p{1})W_2(\p{1}-\p{0})\,
\op{X}^{(02)}(\p{}|\p{1}|\p{0})\no \\  &\odot
\left[\op{X}^{(02)}(\p{}|\p{1}|\p{0})+\op{X}^{(02)}(\p{}|\p{}+\p{0}-\p{1}|
\p{0})\right]\label{ich02}\\
\op{I}^{(03-01)}(\p{}|\p{0})&=W_2(\p{}-\p{0})\left[\op{X}^{(01)}(\p{}|\p{0})
\odot\op{X}^{(03)}(\p{}|\p{0})
+\op{X}^{(03)}(\p{}|\p{0})\odot
\op{X}^{(01)}(\p{}|\p{0})\right]\label{ich03}\, ,
\end{align}
\begin{align}
\op{I}^{(12-10)}(\p{}|\p{0})&=W_1(\p{}-\p{0})\left[\op{X}^{(12)}(\p{}|\p{0})
\odot\op{X}^{(10)}(\p{}|\p{0})
+\op{X}^{(10)}(\p{}|\p{0})\odot
\op{X}^{(12)}(\p{}|\p{0})\right]\label{ich1210}\, ,\\
\op{I}^{(21-01)}(\p{}|\p{0})&=W_2(\p{}-\p{0})\left[\op{X}^{(21)}(\p{}|\p{0})
\odot\op{X}^{(01)}(\p{}|\p{0})
+\op{X}^{(01)}(\p{}|\p{0})\odot
\op{X}^{(21)}(\p{}|\p{0})\right]\label{ich2101}\, ,\\
\op{I}^{(11-11)}(\p{}|\p{0})&= \intp{1}\left[\,W_1(\p{}-\p{1})\,
W_2(\p{1}-\p{0})\,
\op{X}^{(11)12}(\p{}|\p{1}|\p{0}) \right. \no \\ &\odot
\left(\op{X}^{(11)12}(\p{}|\p{1}|\p{0})+\op{X}^{(11)21}(\p{}|\p{}+\p{0}-\p{1}|
\p{0})\right)\no\\
&+\,W_2(\p{}-\p{1})\,W_1(\p{1}-\p{0})\,
\op{X}^{(11)21}(\p{}|\p{1}|\p{0})\no \\ &\odot
\left.\left(\op{X}^{(11)21}(\p{}|\p{1}|\p{0})+\op{X}^{(11)12}(\p{}|\p{}
+\p{0}-\p{1}|
\p{0})\right)\right]\label{ich11}
\end{align}
with 
\begin{align}
\op{X}^{(30)}(\p{}|\p{0})= \intp{1}\,&\left[W_1(\p{1}-\p{0})\,
\op{X}^{(30)}(\p{}|\p{0}|\p{1}|\p{0}) \right. \no \\
&\left.+W_1(\p{}-\p{1})\,(\op{X}^{(30)}(\p{}|\p{1}|\p{0}-\p{}+
\p{1}|\p{0})+\op{X}^{(30)}(\p{}|\p{1}|\p{}|\p{0}))\right]\, ,\label{ch30}\\
\op{X}^{(03)}(\p{}|\p{0})= \intp{1}\,&\left[W_2(\p{1}-\p{0})\,
\op{X}^{(03)}(\p{}|\p{0}|\p{1}|\p{0}) \right. \no \\
&\left.+W_2(\p{}-\p{1})\,(\op{X}^{(03)}(\p{}|\p{1}|\p{0}-\p{}+
\p{1}|\p{0})+\op{X}^{(03)}(\p{}|\p{1}|\p{}|\p{0}))\right]\, ,
\end{align}
\begin{align}
\op{X}^{(12)}(\p{}|\p{0})= \intp{1}\,&\left[W_2(\p{}-\p{1})\left(
\op{X}^{(12)221}(\p{}|\p{1}|\p{}|\p{0})+\op{X}^{(12)212}(\p{}|
\p{1}|\p{0}-\p{}+
\p{1}|\p{0})\right)\right.\no \\ &\left.+W_2(\p{1}-\p{0})\,
\op{X}^{(12)122}(\p{}|\p{0}|\p{1}|\p{0})\right]\, ,\\
\op{X}^{(21)}(\p{}|\p{0})= \intp{1}\,&\left[W_1(\p{}-\p{1})\left(
\op{X}^{(21)112}(\p{}|\p{1}|\p{}|\p{0})+\op{X}^{(21)121}(\p{}|\p{1}|\p{0}-\p{}
+\p{1}|\p{0})\right)\right.\no\\ &\left.+W_1(\p{1}-\p{0})\,
\op{X}^{(21)211}(\p{}|\p{0}|\p{1}|\p{0})\right]\, ,\label{ch21}
\end{align}

\section{Perturbative development and reduced Rayleigh equations}
\label{Perturbative}
In order to obtain the development \eref{dev} for the scattering matrix, 
a practical method is
provided by the use of the reduced Rayleigh equations first obtained by 
Brown {\it et al}
~\cite{Celli} for a single rough surface, and then extended in
\cite{Soubret} for a more general system. These equations are exact, 
under the Rayleigh hypothesis, and their main advantages are that one of the
electric fields  $\E^i$, $\E^s$, $\E^{1-}$, $\E^{1+}$ 
(see \Fref{systeme}) of the problem has been eliminated. 
These equations are derived by taking  linear combinations of the
electromagnetic boundaries conditions at the first surface $h_1$, where the
Fourier transform of the fields has been introduced. 
In particular, we obtain the two
following equations (see Eqs.(99-100) in \cite{Soubret}):
\begin{align}
\intp{} \op{M}_h^{1+,0+}(\vec{u}|\p{})\cdot\op{R}(\p{}|\p{0})\cdot
\E^{i}(\p{0})+
\op{M}_h^{1+,0-}(\vec{u}|\p{0})\cdot\E^{i}(\p{0})=\frac{2\,
(\ep_0\,\ep_1)^{\frac{1}{2}}\,\alu{1}{}}{(\ep_1-\ep_0)}\,
\E^{1+}(\vec{u}) \, ,\label{Ray1}\\
\intp{}\op{M}_h^{1-,0+}(\vec{u}|\p{})\cdot\op{R}(\p{}|\p{0})\cdot
\E^{i}(\p{0})+
\op{M}_h^{1-,0-}(\vec{u}|\p{0})\cdot\E^{i}(\p{0})=-\frac{2\,
(\ep_0\,\ep_1)^{\frac{1}{2}}\,\alu{1}{}}{(\ep_1-\ep_0)}\,
\E^{1-}(\vec{u}) \label{Ray2} \, ,
\end{align}  
where 
\begin{align}
\op{M}_h^{1b,0a}(\vec{u}|\p{})&\equiv\frac{I(b\alu{1}{}-a\alp{0}{}|
\vec{u}-\p{})}{b\alu{1}{}-a\alp{0}{}}\op{M}^{1b,0a}(\vec{u}|\p{}) 
\end{align}
with
\begin{equation}
\op{M}^{1b,0a}(\vec{u}|\p{})= \begin{pmatrix}
||\vec{u}||||\p{}||+ab\,\alu{1}{}\,\alp{0}{}\,\hvec{u}\cdot\hvec{p} &
~~~-b\,\ep_0^{\frac{1}{2}}\,K_0\,\alu{1}{}\,(\hvec{u}\times\hvec{p})_z \\
a\,\ep_1^{\frac{1}{2}}\,K_0\,\alp{0}{}\,(\hvec{u}\times\hvec{p})_z &
(\ep_0\,\ep_1)^{\frac{1}{2}}\,K_0^2\,\hvec{u}\cdot\hvec{p}
\end{pmatrix}\, ,
\label{M10}
\end{equation}
and 
\begin{align}
I(\alpha|\p{})\equiv\int \rmd^2\x\, \e(-\rmi
\p{}\cdot\x-i\alpha\,\hux)\, ,
\label{ialp}
\end{align}
the symbols  $a=\pm$ and $b=\pm$ in \Mref{M10}{ialp} represent a given 
choice linked to the field propagation.
In order to obtain a single equation  for  $\op{R}(\p{}|\p{0})$, we
have to find a relation between $\E^{1-}$ and $\E^{1+}$. To this end
we already know an expression of the 
scattering matrix for a single rough surface separating two homogenous
media of permittivity $\ep_1$ and $\ep_2$, which is translated along 
the $z$-axis to the height $z=-H$, and illuminated by a plane wave 
$E^{1-}(\p{})$. This
scattering  matrix denoted $\op{R}^{H}_{s\,\ep_1,\ep_2}$ is given by:
\begin{equation}
\op{R}^{H}_{s\,\ep_1,\ep_2}(\p{}|\p{0})=\e(\rmi(\alp{1}{}+\alp{1}{0})\,H)\,
\op{R}_{s\,\ep_1,\ep_2}(\p{}|\p{0})\, ,\label{RsH}
\end{equation}
where  $\op{R}_{s\,\ep_1,\ep_2}(\p{}|\p{0})$
can be found in
Refs.~\cite{McGurn,Johnson,Soubret}. The phase term in \Eref{RsH} comes from
the translation $z=-H$ of the rough surface $h_2(\x)$ 
(see Refs.\cite{Voro,Soubret}).
Thus we have the following relation:
\begin{align}
\E^{1+}(\vec{u})=\intu{1}\op{R}^{H}_{s\,\ep_1,\ep_2}(\uu{}|\uu{1})
\cdot\E^{1-}(\vec{u}_1)\, .
\label{E1pm}
\end{align}
Now combining \Eref{E1pm} with \Mref{Ray1}{Ray2}, we obtain an integral 
equation where $\op{R}(\p{}|\p{0})$ is the only unknown:
\begin{align}
\intp{}\,&
\left[\op{M}_h^{1+,0+}(\vec{u}|\p{})+
\intu{1}\,\frac{\alu{1}{}}{\alu{1}{1}}\,\op{R}^{H}_{s\,\ep_1,\ep_2}(\uu{}|
\uu{1})\cdot\op{M}_h^{1-,0+}(\vec{u}_1|\p{})\right]\cdot\op{R}(\p{}|\p{0})=
\no \\
&-\left[\op{M}_h^{1+,0-}(\vec{u}|\p{0})+\intu{1}\,\frac{\alu{1}{}}
{\alu{1}{1}}\,\op{R}^{H}_{s\,\ep_1,\ep_2}(\uu{}|\uu{1})\cdot 
\op{M}_h^{1-,0-}(\vec{u}|\p{0})\right]\, .
\end{align}
Expanding $I(\alpha|\p{})$ in \Eref{ialp} in power of $h_1$:
\begin{align}
I(\alpha|\p{})&=(2\pi)^2\,\delta(\p{})-i\alpha\,h_1^{(1)}(\p{})-
\frac{\alpha^2}{2}\,h_1^{(2)}(\p{})-\frac{i\alpha^3}{3!}\,
h_1^{(3)}(\p{})+\cdots \, ,\label{taylor}
\\
h_1^{(n)}(\p{})&\equiv\int \rmd^2 \x\,\e(-i\p{}\cdot\x)\,h_1^n(\x) \, ,
\end{align}
and using the perturbative development\cite{Soubret} of 
$\op{R}^{H}_{s\,\ep_1,\ep_2}$ 
in power of $h_2$:
\begin{eqnarray}
\op{R}^{H}_{s\,\ep_1,\ep_2}(\p{}|\p{0})&=&(2\pi)^2\delta(\p{}-\p{0})\,
\op{X}^{H(0)}_{s\,\ep_1,\ep_2}(\p{0})+
\alp{0}{0}\,
\op{X}^{H(1)}_{s\,\ep_1,\ep_2}(\p{}|\p{0})\, h_2(\p{}-\p{0}) 
\label{Dev1}\nonumber\\
& & +\alp{0}{0}\,
\intp{1}\,\op{X}^{H(2)}_{s\,\ep_1,\ep_2}(\p{}|\p{1}|\p{0})\, h_2(\p{}-\p{1})
h_2(\p{1}-\p{0}) \nonumber \\ 
& &+\alp{0}{0}\,
\iintp{1}{2} \,\op{X}^{H(3)}_{s\,\ep_1,\ep_2}(\p{}|\p{1}|
\p{2}|\p{0})\, h_2(\p{}-\p{1})
h_2(\p{1}-\p{2})h_2(\p{2}-\p{0})  \, ,\no \\ \label{Dev2}
\end{eqnarray}
we finally obtain the expansions \esref{dev}{R10}{R03}.
The expression for the scattering matrix when only 
one rough surface is involved was given in
Ref.\cite{Soubret}, here a small change have been made in the notations:
\begin{align}
\op{R}^{(n0)}(\p{}|\p{0})=\op{R}_u^{(n)}(\p{}|\p{0})\, ,\\
\op{R}^{(0n)}(\p{}|\p{0})=\op{R}_d^{(n)}(\p{}|\p{0})
\end{align}
$n$ being an integer ranging from $0$ to $3$.
For the others coefficients we have :
\begin{align}
\op{X}^{(11)21}(\p{}|\p{1}|\p{0})=&\op{T}^{10}(\p{})
\cdot\op{U}^{(0)}(\p{})\cdot\op{X}^{H(1)}_{s\,\ep_1,\ep_2}(\p{}|\p{1})
\cdot\left[-\op{\ep}\cdot\op{D}_{10}^{\,-}(\p{1})
\cdot\op{X}^{(10)}(\p{1}|\p{0})+i\op{S}^+(\p{1}|\p{0})\right]\, 
,\\
\op{X}^{(11)12}(\p{}|\p{1}|\p{0})=&\rmi\,
\op{P}^{+}(\p{}|\p{1})\cdot\op{X}^{(01)}(\p{1}|\p{0})\, ,\\
\op{X}^{(21)112}(\p{}|\p{1}|\p{2}|\p{0})=&\rmi\,
\op{P}^{+}(\p{}|\p{1})\cdot\op{X}^{(11)12}(\p{1}|\p{2}|\p{0})\no \\
&+\frac{1}{2}\,\left[\alp{1}{}\,\op{P}^{\,-}(\p{}|\p{2})-\alp{0}{2}\,
\op{P}^+(\p{}|\p{2})\right]\cdot 
\op{X}^{(01)}(\p{2}|\p{0})\,\\
\op{X}^{(21)121}(\p{}|\p{1}|\p{2}|\p{0})=&\rmi\,
\op{P}^{+}(\p{}|\p{1})\cdot\op{X}^{(11)21}(\p{1}|\p{2}|\p{0})\,,\\
\op{X}^{(21)211}(\p{}|\p{1}|\p{2}|\p{0})=&\op{T}^{10}(\p{})\cdot
\op{U}^{(0)}(\p{})\cdot\op{X}^{H(1)}_{s\,\ep_1,\ep_2}(\p{}|\p{1})
\cdot\left[-\op{\ep}\cdot\op{D}_{10}^{\,-}(\p{1})\cdot
\op{X}^{(20)}(\p{1}|\p{2}|\p{0})\right.\no\\
&+\frac{i\,(\ep_1-\ep_0)}{2\,(\ep_0\,\ep_1)^{1/2}}\op{M}^{1-,0+}(\p{1}|\p{2})
\cdot\op{X}^{(10)}(\p{2}|\p{0})\no 
\\ &\left.-\frac{1}{2}\left(\alp{1}{1}\,\op{S}^+(\p{1}|\p{0})+\alp{0}{0}\,
\op{S}^{\,-}(\p{1}|\p{0})\right)\right]\,\\
\op{X}^{(12)122}(\p{}|\p{1}|\p{2}|\p{0})=&\rmi\,
\op{P}^{+}(\p{}|\p{1})\cdot\op{X}^{(02)}(\p{1}|\p{2}|\p{0})\,,\\
\op{X}^{(12)212}(\p{}|\p{1}|\p{2}|\p{0})=&\op{T}^{10}(\p{})
\cdot\op{U}^{(0)}(\p{})\cdot\op{X}^{H(1)}_{s\,\ep_1,\ep_2}(\p{}|
\p{1})\cdot\left[-\op{\ep}\cdot\op{D}_{10}^{\,-}(\p{1})\cdot
\op{X}^{(11)12}(\p{1}|\p{2}|\p{0})\right.\no\\
&\left.+\frac{i\,(\ep_1-\ep_0)}{2\,(\ep_0\,\ep_1)^{1/2}}\op{M}^{1-,0+}(\p{1}|
\p{2})\cdot\op{X}^{(01)}(\p{2}|\p{0})\right]\, 
,\\
\op{X}^{(12)221}(\p{}|\p{1}|\p{2}|\p{0})=&\op{T}^{10}(\p{})\cdot
\op{U}^{(0)}(\p{})\cdot\left[-\op{X}^{H(1)}_{s\,\ep_1,\ep_2}(\p{}|\p{1})
\cdot\op{\ep}\cdot\op{D}_{10}^{\,-}(\p{1})\cdot\op{X}^{(11)21}(\p{1}|
\p{2}|\p{0})\right.\no\\
&\left.+\,\op{X}^{H(2)}_{s\,\ep_1,\ep_2}(\p{}|\p{1}|\p{2})
\cdot\left(-\op{\ep}\cdot\op{D}_{10}^{-}(\p{2})\cdot\op{X}^{(10)}(\p{2}|
\p{0})+\rmi\,\op{S}^+(\p{2}|\p{0})\right)\right]\,,\\
\end{align}
where the matrices $\op{T}^{(10)}$, $\op{U}^{(0)}$, $\op{P}^{\pm}$,
$\op{D}_{10}^{\pm}$, $\op{X}^{(10)}$, $\op{X}^{(20)}$, $\op{X}^{(01)}$,  
$\op{X}^{(02)}$,  $\op{X}^{H\,(1)}_{s\,\ep_1,\ep_2}$, 
$\op{X}^{H\,(2)}_{s\,\ep_1,\ep_2}$ are respectively defined by 
Eqs.(71),(87),(114,115),(57),(104,105),(96,97),(60,61) in 
Ref.~\cite{Soubret}, and $\op{\ep}$, $\op{S}^{\pm}$ are given in 
Appendix \ref{appA}.
\section{Numerical results}
\label{Numerical}
\begin{figure}[ht]
\input{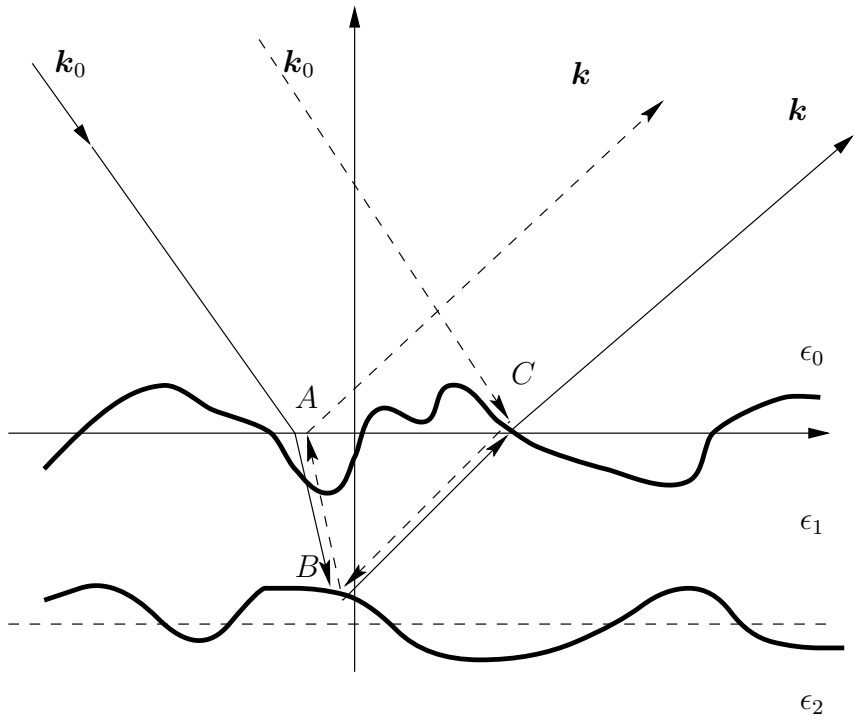}
\caption{New mechanisms responsible of enhanced backscattering. The
  incident wave is transmitted in $A$ as perfect interface between the media
  $0$ and $1$, then the wave is scattered by the second rough
  surface in $B$ and by the first one in $C$. But the wave can 
  now  follow
  this path the other way round. The phase difference between
  these two waves is zero in the anti-specular direction which produces
  the peak.}
\label{systeme0}
\end{figure}

\begin{figure}[ht]
\includegraphics{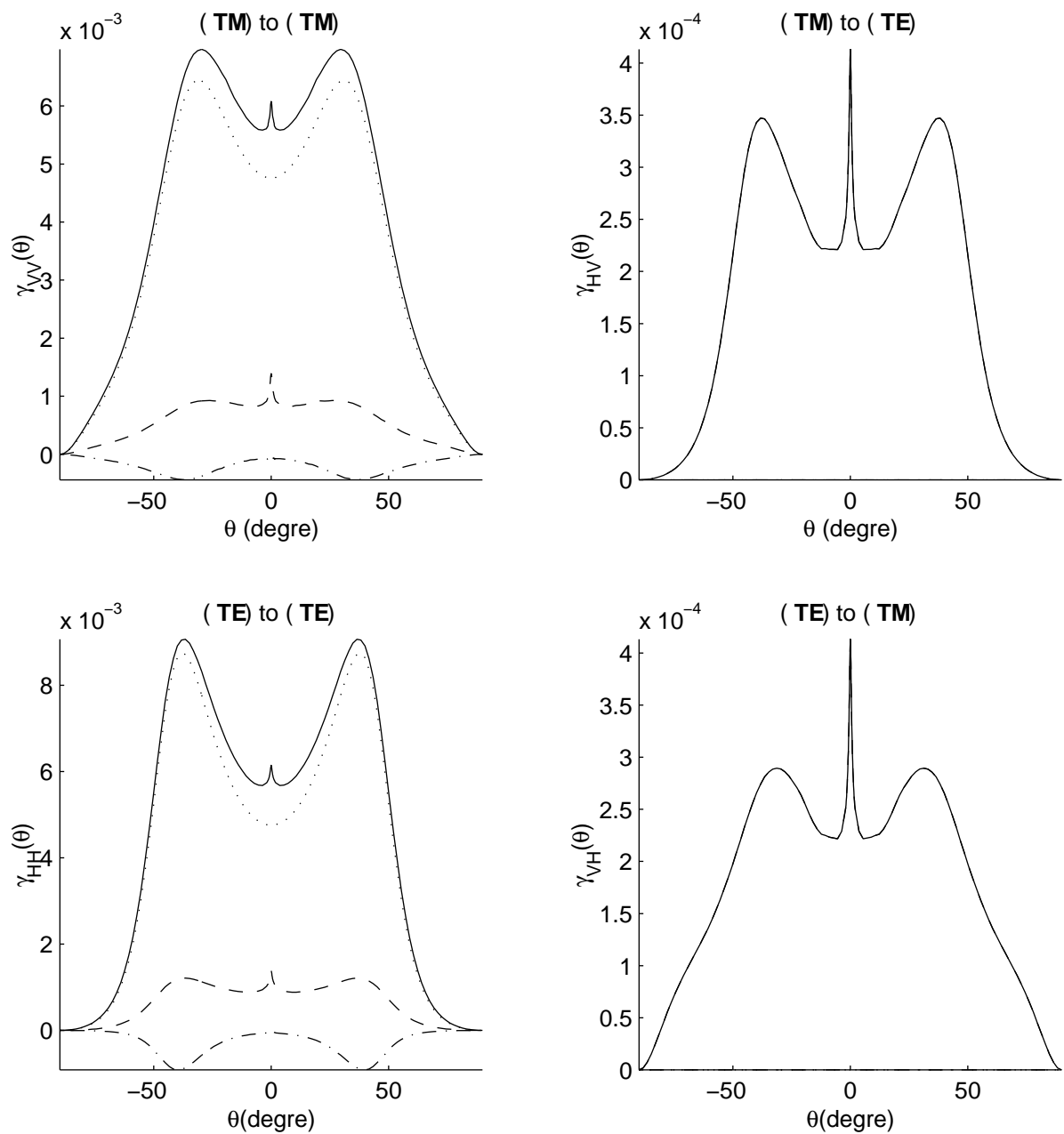}
\caption{The bistatic coefficients for an horizontal $(TE)$ and a vertical 
$(TM)$ polarized normally incident light of wavelength $\lambda=632.8\,nm$,
on a slab
with an upper two-dimensional randomly rough surface, characterized by the
parameters $\sigma_1=15 \,nm$, $l_1=100\,nm$, and a bottom rough
surface   characterized by $\sigma_2=5 \,nm$, $l_2=100\,nm$. The
dielectric constants are $\epsilon_1=2.6896+i\,0.0075$ and 
$\epsilon_2=-18.3+i\,0.55$.
The thickness of the film is $H=500\,nm$. 
The scattered field is observed in the incident plane. 
For each figure are plotted : the total incoherent
scattering $\op{\ga}^{incoh}$ (solid curve), the first order given by
$\op{I}^{(10-10)}+\op{I}^{(01-01)}$ (dotted curve), 
the second order $\op{I}^{(20-20)}+\op{I}^{(02-02)}+\op{I}^{(11-11)}$ 
(dashed curve), and the
third order $\op{I}^{(30-10)}+\op{I}^{(03-01)}+\op{I}^{(12-10)}+
\op{I}^{(21-01)}$ (dash-dotted curve).}
\label{Normale}
\end{figure}

\begin{figure}[ht]
\includegraphics{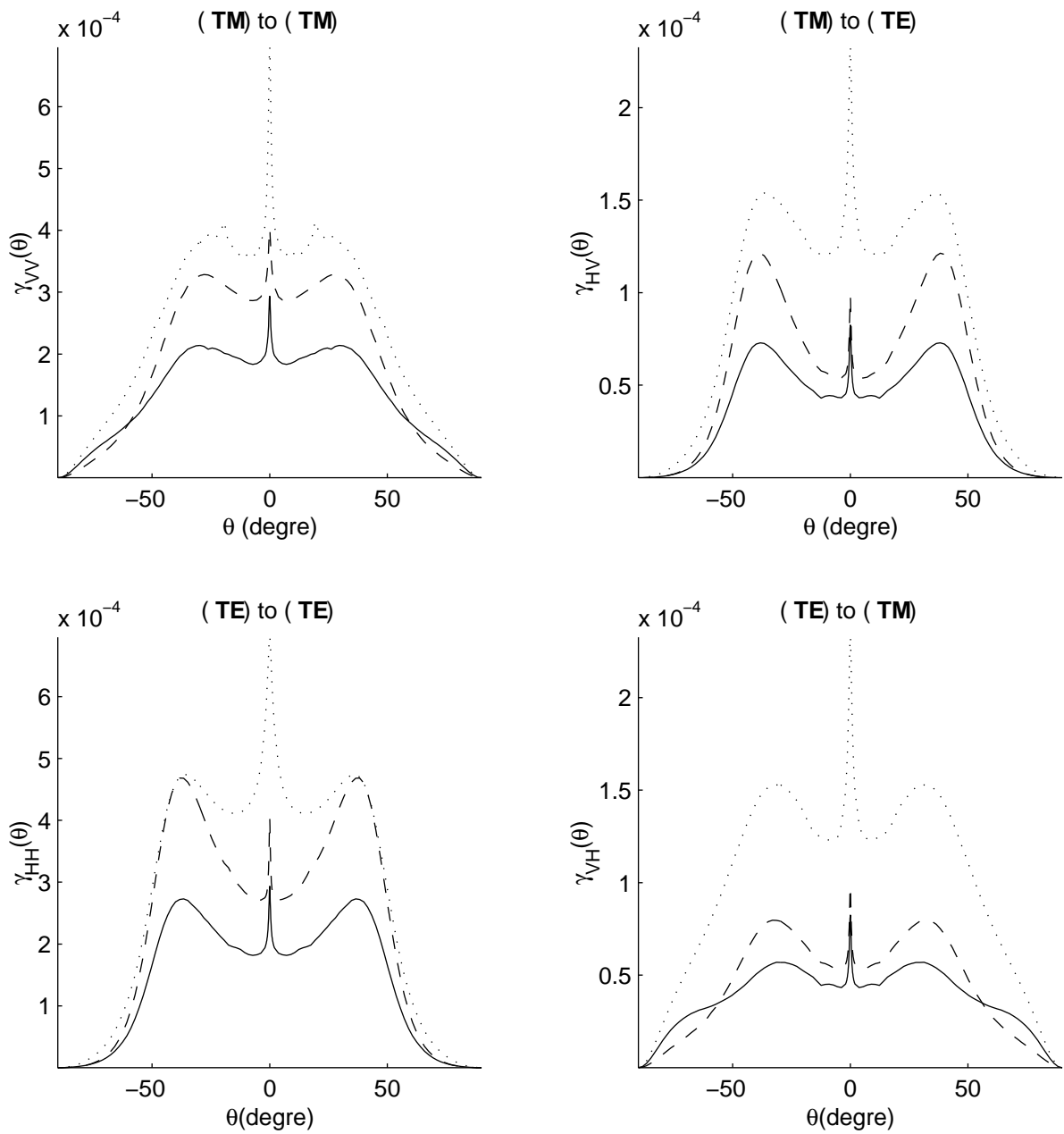}
\caption{Only the second order contributions of \Fref{Normale} are
  depicted. $\op{I}^{(11-11)}$ is the solid curve, $\op{I}^{(20-20)}$
  the dashed curve and $\op{I}^{(02-02)}$ the dotted curve.}
\label{Normale22}
\end{figure}

\begin{figure}[ht]
\includegraphics{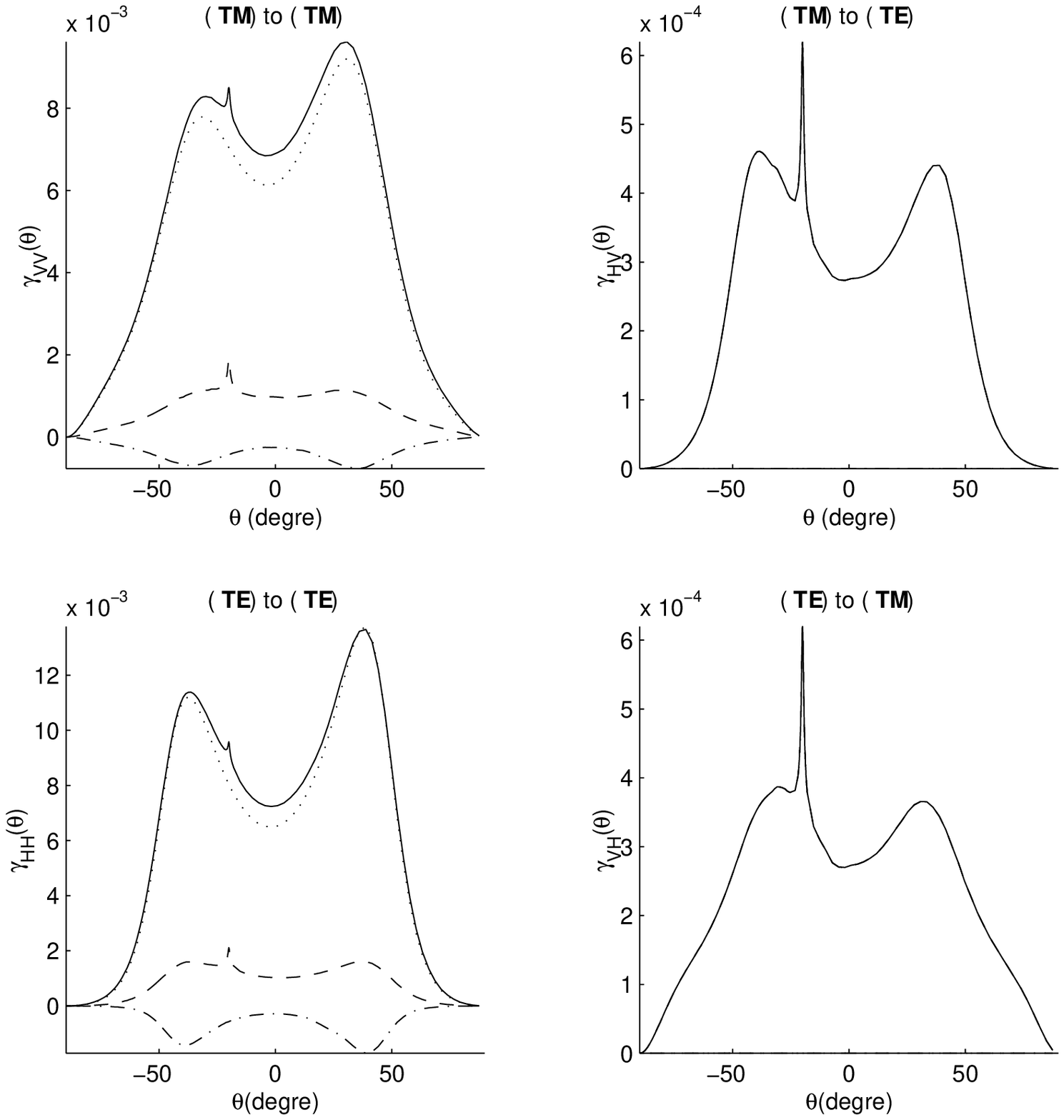}
\caption{Same parameters as in \Fref{Normale}, but with $\theta_0=-20\degre$}
\label{Inc20}
\end{figure}

\begin{figure}[ht]
\includegraphics{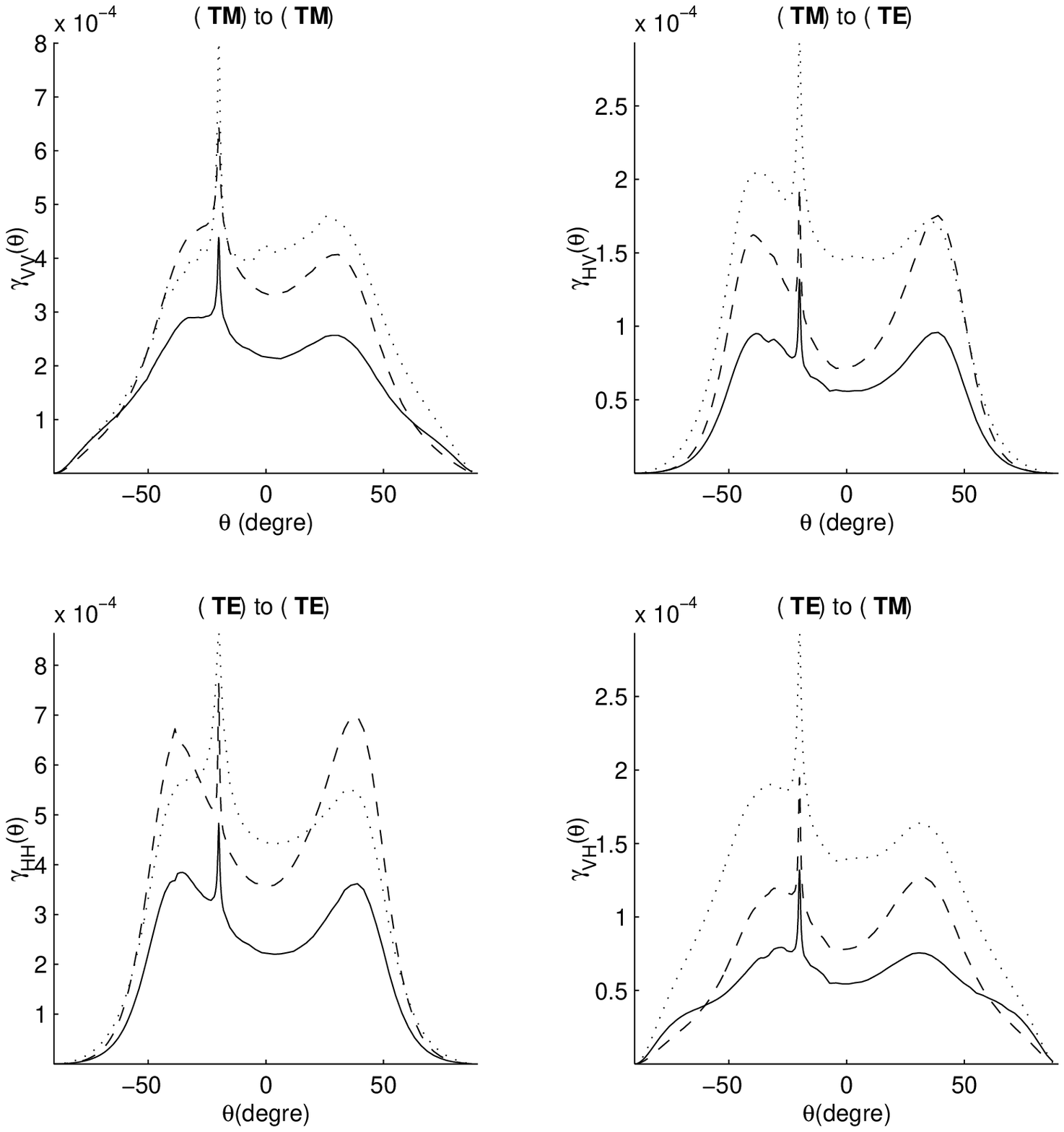}
\caption{Second order contributions to \Fref{Inc20}}
\label{Inc2022}
\end{figure}

As an application of the previous formalism we consider a system made of
a air-dielectric film whose dielectric constant is 
$\ep_1=2.6896+i\,0.0075$($\ep_0=1$), 
deposited on silver surface with $\ep_2=-18.3+i\,0.55$. The
vacuum-dielectric interface is a two-dimensionnal rough surface, whose 
parameters are $\sigma_1=15\,nm$ and $l_1=100\,nm$. The
dielectric-silver boundary is also rough  and defined by
$\sigma_2=5\,nm$ and $l_2=100\,nm$. The incident wave has an arbitrary 
polarization and his wavelength is $\lambda=632.8\,nm$. With this set 
 of parameters, the conditions of validity of the
small-perturbation  theory are satisfied \cite{Ogilvy} namely:
\begin{alignat}{2}
2\pi\,\left|\frac{\ep_1}{\ep_0}\right|^{1/2}\,\frac{\sigma_1}{\lambda}&\ll1
\, ,&
\qquad
2\pi\,\left|\frac{\ep_2}{\ep_1}\right|^{1/2}\,\frac{\sigma_2}{\lambda}&\ll1\\
\frac{\sigma_1}{l_1}&\ll1 \,, & \qquad  \frac{\sigma_2}{l_2}&\ll1\,.
\end{alignat}
The thickness of the film is $H=500\,nm$ and support two-guided wave
polaritons 
for the $(TE)$ polarizations at $p_{TE}^{1}=1.5534\,K_0$, and 
$p_{TE}^{2}=1.2727\,K_0$, and three guided-modes for the $(TM)$
polarizations at $p_{TM}^{1}=1.7752\,K_0$, $p_{TM}^{2}=1.4577\,K_0$ and 
$p_{TM}^{3}=1.034\,K_0$.
We have computed the incoherent bistatic coefficient $\ga^{incoh}(\p{}|\p{0})$, 
where
the integrals involved in Eqs.~(\ref{ich20},\ref{ich02},\ref{ich11},
\ref{ch30}-\ref{ch21}) are 
evaluated using Legendre quadrature. The results are shown in
Figs.~(\ref{Normale}-\ref{Inc2022}), where the incoherent bistatic
coefficient is drawn as of function of the scattering angle $\theta$
for two different angles of incidence and the incident wave is
linearly polarized. In \Fref{Normale}, the wave is normally incident
and the scattered field is observed in the incident plane
($\phi=0\degre$).
The single scattering contribution on each surface, associated with the 
terms $\op{I}^{(10-10)}+\op{I}^{(01-01)}$, is plotted as a dotted line,
the double-scattering contribution
$\op{I}^{(20-20)}+\op{I}^{(02-02)}+\op{I}^{(11-11)}$ as a dashed line, 
the other terms
$\op{I}^{(30-10)}+\op{I}^{(03-01)}+\op{I}^{(12-10)}+\op{I}^{(21-01)}$
as a dash-dotted line, and the total contribution
$\ga^{incoh}$ by the solid curve.  We observed an enhancement of the
backscattering which corresponds to the physical process in which the
incident light excites a guided-mode through the roughness of the slab and 
then  is
scattered into a volume wave  which is also due to the roughness effect. 
During the same process,the light can follow this path in the opposite 
direction where one possible configuration is
shown in \Fref{systeme0}. These two paths can interfere constructively 
near the backscattering direction to produce a peak. As these paths are
identical for the two waves under consideration, they should have the same 
degree of
interaction with the rough surface, thus a term like $\op{I}^{(30-10)}$
cannot produce the peak because the first wave interact three times
with the upper rough surface while the second wave only one time. The effect 
comes only from the terms  $\op{I}^{(20-20)}+\op{I}^{(02-02)}+\op{I}^{(11-11)}$,
which contain the paths indicated in  \Fref{systeme0}. However, it has 
to be noticed that these terms contain also paths that do not produce
enhanced backscattering, for example in  $\op{I}^{(20-20)}$ we
have the scattering process where the incident wave is only
scattered one time by the upper rough surface but the scattering
process is of order two in $h_1$. This is the reason why the terms
$\op{I}^{(20-20)}+\op{I}^{(02-02)}+\op{I}^{(11-11)}$ are not zero away
from the anti-specular direction. In order to separate the
different contributions to the  backscattering peak, we have drawn the
contributions of $\op{I}^{(20-20)}$, $\op{I}^{(02-02)}$,
$\op{I}^{(11-11)}$ separately in \Fref{Normale22} as a dashed-line,
dotted-line and solid curve respectively, we see that each term
produces an enhancement near the anti-specular direction. The terms
$\op{I}^{(20-20)}$, $\op{I}^{(02-02)}$ are the classic
one(see Refs.~\cite{Mara,Soubret}) where the fields do not interact
with both rough surfaces but produce the peak due to
the scattering on the same rough surface: the upper one for
$\op{I}^{(20-20)}$ and the bottom one for $\op{I}^{(02-02)}$.
The contribution $\op{I}^{(11-11)}$ is {\it the new result} of this
work, because we see that the mechanism of \Fref{systeme0} exists and has
 the same magnitude as the other terms for the  choosen parameters.
In order to clearly show the displacement of the backscattering enhancement
as the angle of incidence is varied, we show in
Figs. \ref{Inc20} and \ref{Inc2022} the numerical results of the
perturbation method when $\theta_0=20\degre$. As expected, we clearly 
see the peak which is now located at $\theta_0=-20\degre$.
Recent papers \cite{Mara,Soubret} have also explored the satellite
peaks phenomenon which occurs when the wave follows  two reverse paths but
with different guided-mode excitations. In \Fref{Normale22}, this phenomenon 
appears for the $(TM)$ to $(TM)$ polarization due to  the term
$\op{I}^{(02-02)}$. Although a similar phenomenon occurs for the other terms 
it is too weak to contribute significantly..
\section{Conclusion}
\label{Conclusions}
In conclusion, the results in this paper clearly show that the
backscattering enhancement produced by a rough slab due to the
guided-wave polaritons has several origin. At small order in the
perturbative development, the peak is produced by a double scattering
mechanism. The new results we would like to emphasize, is that not only
processes where the two scattering events take place on the same rough
surface are capable to produce the backscattering peak, but it is also due to 
the fact that the wave can be scattered first by one of the rough 
surface and second by the other surface. 
To carry the proof, we have performed a perturbative
development up to order four in the rms-height of the surfaces which 
has been possible using an integral equation called reduced Rayleigh
equation, in which  only the scattering matrix of the whole structure is 
unknown.

\section*{Acknowledgments}
(AS) thanks ANRT for financial support during the preparation of
his thesis (contract CIFRE-238-98).

\appendix
\section{Scattering matrix coefficients}
\label{appA}
Here we collect the expression of the matrices not given in
\cite{Soubret}:
\begin{align}
\op{\ep}&\equiv\frac{1}{2}\begin{pmatrix}
(\ep_0\,\ep_1)^{-1/2} & 0 \\
0 & 1
\end{pmatrix}\\
\op{S}^{\pm}(\p{}|\p{0})&\equiv\frac{\ep_1-\ep_0}{2\,\alp{0}{0}\,
(\ep_0\,\ep_1)^{1/2}}\,\left[\op{M}^{1-,0+}(\p{}|\p{0}).
\cdot\op{X}^{(00)}(\p{}|\p{0})\pm\op{M}^{1-,0-}(\p{}|\p{0})\right]\, .
\end{align}
After some calculations we obtain :
\begin{align}
\op{S}^{+}(\p{}|\p{0})=&\frac{(\ep_1-\ep_0)}{(\ep_0\,\ep_1)^{1/2}}\no\\
&\times\begin{pmatrix}
\ep_1\,||\p{}||\,||\p{0}||\,F^+_V(\p{0})+\ep_0\,\alp{1}{}\,\alp{1}{0}\,
F^-_V(\p{0})\,\hvec{p}\cdot\hvec{p}_0 
&
\ep_0^{1/2}\,K_0\,\alp{1}{}\,F^+_H(\p{0})\,(\hvec{p}\times\hvec{p}_0)_z \\
-\ep_0\,\ep_1^{1/2}\,K_0\,\alp{1}{0}\,F_V^-(\p{0}) \,(\hvec{p}\times
\hvec{p}_0)_z
&
(\ep_0\,\ep_1)^{1/2}\,K_0^2\,F_H^+(\p{0})\,\hvec{p}\cdot\hvec{p}_0\end{pmatrix}
\no \\ 
&\cdot[\op{D}_{10}^+(\p{0})]^{-1}\, ,\\
\op{S}^{\,-}(\p{}|\p{0})=&\frac{(\ep_1-\ep_0)}{\alp{0}{0}\,
(\ep_0\,\ep_1)^{1/2}}\no \\
&\times\begin{pmatrix}
-\ep_0\,\alp{1}{0}\,||\p{}||\,||\p{0}||\,F^-_V(\p{0})
&
-\ep_0^{1/2}\,K_0\,\alp{1}{}\,\alp{1}{0}\,F^-_H(\p{0})\,(\hvec{p}\times
\hvec{p}_0)_z \\
-\ep_1\,\alp{1}{}\,\alpha^2_{0}(\p{0})\,F^+_V(\p{0})\,\hvec{p}\cdot\hvec{p}_0
&  \\ &\\
\ep_1^{3/2}\,K_0\,\alpha^2_{0}(\p{0})\,F_V^+(\p{0}) \,(\hvec{p}\times
\hvec{p}_0)_z
&-(\ep_0\,\ep_1)^{1/2}\,K_0^2\,\alp{1}{0}\,F_H^-(\p{0})\,\hvec{p}
\cdot\hvec{p}_0\end{pmatrix}\no\\
&\cdot[\op{D}_{10}^+(\p{0})]^{-1}\, .
\end{align}


\listoffigures
\newpage
\newpage

\end{document}